\documentclass[12pt]{article}
\usepackage{amsfonts}
\usepackage{amsmath, amsthm}
\usepackage{epsfig}
\usepackage{algorithm}
\usepackage{algorithmic}
\usepackage{epic}
\usepackage{graphicx}    
\usepackage{amssymb}     
\usepackage{enumitem}
\usepackage{tabularx}
\usepackage{booktabs}
\usepackage{multirow}
\usepackage{longtable}

\usepackage{verbatim}
\usepackage{color}
\usepackage{bm}
\usepackage{ragged2e}
\usepackage{changepage}
\usepackage{array}
\usepackage[table]{xcolor}
\definecolor{Gray}{gray}{.80}
\usepackage{lscape}
\usepackage{apacite}
\usepackage[round,authoryear]{natbib}

\usepackage[normalem]{ulem}
\usepackage[colorlinks=true,urlcolor=blue,citecolor=purple,linkcolor=blue,bookmarks=true]{hyperref}
\usepackage{caption}
\usepackage{pdfsync}
\usepackage{natbib}

\usepackage{booktabs}  
\usepackage{adjustbox} 
\usepackage{fancyhdr} 

\usepackage{makecell}

\usepackage{subcaption}  
\usepackage{float}       
\usepackage{authblk}
\usepackage{footnotehyper}
\usepackage{threeparttable} 
\usepackage{adjustbox}
\usepackage{siunitx}
\usepackage{setspace}
\usepackage{tikz}
\usetikzlibrary{positioning, shapes.geometric, arrows}

\tikzset{
  block/.style = {
    rectangle, 
    draw, 
    text centered, 
    minimum width=4.5cm,  
    minimum height=1.0cm  
  },
  arrow/.style = {thick,->,>=stealth}
}
\usepackage{xr}
\usepackage{hyperref} 
\usepackage{cleveref} 

\usepackage{titlesec}
\usepackage{booktabs}
\titlelabel{\thetitle.\quad}

\begin{document}
\baselineskip=16pt
\begin{center}
	{\Large \bf Is Repeated Bayesian Interim Analysis Consequence-Free?} 
\end{center}

\begin{center}
	Suyu Liu$^1$, Beibei Guo$^2$,  Laura Thompson$^3$, Lei Nie$^3$, Ying Yuan$^{2, *}$,\\
\end{center}
\noindent{
$^1$Department of Biostatistics, The University of Texas MD Anderson Cancer Center, Houston, TX\\
$^2$Department of Experimental Statistics, Louisiana State University, Baton Rouge, LA \\
$^3$ Office of Biostatistics, OTS/CDER, Food and Drug Administration, Silver Spring, MD\\
$^*$ Email: yyuan@mdanderson.org

\baselineskip=24pt

\begin{center}
\textbf{Abstract}
\end{center}

\vspace{1mm}

Interim analyses are vital in clinical trials for early decision-making. While frequentist implications are well-established, the consequences of repeated Bayesian interim monitoring for efficacy, specifically regarding multiplicity, remain contentious. This article provides theoretical justification and numerical evidence evaluating the impact of such designs on bias, mean squared error (MSE), credible interval coverage, false discovery rate (FDR), and average Type I error (ATIE).
Our findings show that when the inferential prior matches the data-generating prior, sequential efficacy stopping does not bias the posterior mean or degrade credible interval coverage. However, even under this ``matched" condition, the FDR, ATIE, and MSE are significantly altered. In the more practically relevant scenario where the inferential and data-generating priors differ, all aforementioned operating characteristics, including estimation bias and coverage, are substantially impacted. These results reconcile long-standing conflicting arguments regarding Bayesian multiplicity. We demonstrate that while some Bayesian properties are invariant to sequential looks, others are not. Our work underscores the necessity of thoughtful prior specification and comprehensive evaluation of frequentist-Bayesian operating characteristics to ensure reliable inference in adaptive trial designs.

\vspace{5mm}

\section{Introduction}
In clinical trials,  interim analyses are often conducted to assess trial progress and inform decisions about whether to continue or stop the trial early. A trial may be stopped early if interim results cross predefined thresholds, such as showing compelling efficacy, or indicating futility due to lack of benefit. In this article, we focus on interim analyses that may lead to early stopping and claim of efficacy, refer to as superiority stopping. Two primary types of decision rules are used in interim analyses for superiority stopping: frequentist and Bayesian. 

In the frequentist approach, the impact of interim analyses on the operating characteristics of clinical trials, particularly the familywise error rate (FWER), has been extensively studied and is generally understood. Within this framework, parameters are treated as fixed but unknown constants. Repeated interim looks with superiority stopping introduce a multiplicity problem, leading to inflation of the FWER. To address this, various methods have been developed to adjust significance levels or decision rules at each interim analysis, thereby ensuring that the overall FWER remains controlled at the desired level. The Pocock method \citep{Pocock77} assigns equally stringent significance levels to all interim analyses, making early stopping easier but potentially sacrificing efficiency and power. In contrast, the O'Brien-Fleming method \citep{OBrien79} applies more stringent stopping criteria early in the trial and gradually relaxes them as the trial progresses, offering a more conservative approach. Alpha spending functions \citep{Lan83} provide additional flexibility by allocating the total alpha level across multiple analyses and dynamically adjusting critical values based on how much alpha has already been `spent'.

In contrast, the Bayesian framework, which treats parameters as random variables with associated probability distributions, has been less thoroughly examined in the context of interim analyses, and no clear consensus exists regarding their effects on trial operating characteristics. \citet{Berry87} argued that Bayesian interim analyses with superiority stopping are not subject to multiplicity concerns because they adhere to the likelihood principle, allowing analyses to be conducted at any point during the trial without adjustment. \citet{Harrell17} echoed this perspective, advocating the use of Bayesian posterior probabilities to make go/no-go decisions as frequently as needed.

However, this view has been met with skepticism. \citet{Armitage58} noted that multiple interim looks can affect the maximum likelihood estimator and the coverage probability of confidence intervals. \citet{Jennison90} further cautioned that Bayesian methods may exhibit poor frequentist properties, including inflated FWER when multiple interim analyses are conducted. \citet{Rosenbaum84} conducted simulation studies showing that the coverage probability of posterior credible intervals can be distorted when commonly used flat, non-informative priors are employed for inference. Most critiques in the literature focus primarily on frequentist quantities, particularly FWER, raising questions about their relevance and about the appropriateness of using frequentist metrics to assess Bayesian inference. The U.S. Food and Drug Administration has also acknowledged this issue, stating in recent guidance \citep{FDA19} that, relative to FWER inflation, “biased estimation in adaptive design is currently a less well-studied phenomenon.”

In this article, we aim to fill this critical knowledge gap by providing both theoretical and numerical investigations into how repeated Bayesian interim analyses affect the statistical properties of clinical trial designs. Specifically, we examine their impact on the bias and mean squared error (MSE) of the posterior mean, the coverage probability of posterior credible intervals, and error metrics such as the false discovery rate (FDR) and average type I error (ATIE)—quantities that, while often viewed through a frequentist lens, are evaluated here within a fully Bayesian framework. Our analysis explicitly distinguishes between the data-generating prior and the user-specified prior, accounting for the probability distribution of unknown parameters to maintain a strictly Bayesian perspective.

While some argue that Bayesian interim monitoring allows for continuous looks without adjustment, we show this claim holds only for specific quantities, such as the posterior mean and credible interval coverage. More importantly, it relies on the assumption that the user-specified (i.e., analyst’s) prior perfectly matches the true data-generating prior. This ``matched-prior" assumption is rarely met in practice and has been largely unacknowledged in seminal studies, including \citet{Berry87}. By identifying these boundary conditions, our findings reconcile the conflicting arguments regarding Bayesian multiplicity that have long persisted in the literature.

Moreover, we demonstrate that even when the two priors coincide, repeated Bayesian interim analyses can systematically distort other key statistical properties, including the FDR and ATIE. Crucially, when the user-specified prior deviates from the data-generating prior, the typical scenario in practice, all evaluated metrics are adversely affected.

The remainder of this article is organized as follows. In Section 2, we present Bayesian probability models for binary, normal, and time-to-event outcomes, along with a theoretical assessment of the impact of repeated Bayesian interim analyses on the operating characteristics of clinical trials. Section 3 evaluates these effects through extensive simulation studies. We offer concluding remarks in Section 4.

\section{Methods}
\subsection{Probability Models}
Consider a superiority trial that aims to demonstrate the superiority of an experimental treatment over a control,  with interim analyses that may lead to early termination if there is overwhelming evidence of superiority. Let $X_i$ denote the efficacy endpoint for the $i$th patient receiving the experimental treatment, where $X_i$ follows a probability distribution indexed by the parameter $\theta$, representing the treatment effect of interest,  possibly along with other nuisance parameters. Without loss of generality, we assume that larger values of $\theta$ indicate higher efficacy.

Next, we present commonly used Bayesian models for binary, normal, and time-to-event endpoints, along with their associated prior and posterior distributions. For binary and normal outcomes in randomized controlled trials (RCTs), patients in the treatment and control arms are independent, allowing the same model to be applied separately to each arm. As a result, we do not describe separate models for single-arm trials and RCTs in these cases. For time-to-event endpoints, considering the popularity of using the Cox proportional hazards model \citep{Cox72}  in RCTs, we present separate modeling approaches for single-arm trials and RCTs. These models are not required for the theoretical development presented in the next section, but they facilitate the understanding of theoretical results and serve as the foundation for the simulation study.

\noindent\textit{Binary endpoint:} For a binary endpoint $X_i$, such as tumor response, we assume
\begin{equation*}
X_i\sim \mbox{Bernoulli}(\theta),
\end{equation*}
where $\theta$ denotes the response rate. Suppose the user-specified prior for $\theta$ is $\pi(\theta) =  \mbox{Beta}(\alpha,\beta)$, where $\alpha$ and $\beta$ are hyperparameters. Given interim data ${\cal D}_n=(X_1,\cdots,X_n)$ from the first $n$ patients in the trial, the posterior distribution for $\theta$ under the user-specified prior is 
$$\pi(\theta|{\cal D}_n) = \mbox{Beta}(\alpha+\sum_{i=1}^nX_i, \beta+n-\sum_{i=1}^nX_i)$$ with posterior mean $\hat{\theta} = \frac{\alpha+\sum_{i=1}^nX_i}{\alpha+\beta+n}$.

\noindent\textit{Normal endpoint:} For a continuous outcome, we assume
\begin{equation*}
X_i\sim N(\theta,\sigma^2),
\end{equation*}
where $\theta$ and $\sigma^2$ represent the mean and variance, respectively. Suppose the user-specified joint prior distribution of $\theta$ and $\sigma^2$ is a conjugate prior $\pi(\theta,\sigma^2)=\pi(\theta|\sigma^2)\pi(\sigma^2)$, with $\pi(\theta|\sigma^2)=N(\mu,\sigma^2/\kappa)$ and $\pi(\sigma^2)=\mbox{Inv-}\chi^2(\nu,\sigma_0^2)$, where $\mu$, $\kappa$, $\nu$, and $\sigma_0^2$ are hyperparameters. Here, $\nu$ and $\sigma_0^2$ denote the degrees of freedom and scale parameter for the inverse chi-squared distribution. Given interim data ${\cal D}_n=(X_1,\cdots,X_n)$ from the first $n$ patients with sample mean $\bar{X}_n=(X_1+\cdots +X_n)/{n}$, the marginal posterior distribution of $\theta$ under the user-specified prior is $$\pi(\theta|{\cal D}_n)= t_{\nu+n}(\nu_n,\tau_n),$$ where the location (also the posterior mean, if $\nu+n>1$)  and scale parameters are given by $$\nu_n=\frac{n\bar{X}_n+\kappa\mu}{\kappa+n}, \quad \tau_n=\big\{\nu\sigma_0^2+\sum(X_i-\bar{X}_n)^2+\frac{\kappa n}{\kappa+n}(\bar{X}_n-\mu)^2\big\}/(\nu+n)(\kappa+n).$$

In order to focus on the effect of the prior distribution of $\theta$ on interim analyses, we also consider the case where the variance $\sigma^2$ is known. Specifically, we assume the data $X_i$ is standardized and set $\sigma^2 = 1$. In this setting, the user-specified prior for $\theta$ is the conjugate prior $\theta\sim N(\mu,\sigma_0^2)$, with $\mu$ and $\sigma_0^2$ as hyperparameters. Given interim data ${\cal D}_n$ with sample mean $\bar{X}_n$, the posterior distribution for $\theta$ under the user-specified prior is
$\pi(\theta|{\cal D}_n)= N(\mu_n,\sigma_n^2)$, 
where the posterior mean $\mu_n$ and variance $\sigma_n^2$ are given by
\begin{equation*}
\mu_n=\frac{\frac{1}{\sigma_0^2}\mu+\frac{n}{\sigma^2}\bar{X}_n}{\frac{1}{\sigma_0^2}+\frac{n}{\sigma^2}}\ \ \ \ \ \ \mbox{and}\ \ \ \ \ \ \frac{1}{\sigma_n^2}=\frac{1}{\sigma_0^2}+\frac{n}{\sigma^2}.
\end{equation*}

\noindent\textit{Single-arm trials with time-to-event endpoint:} For a time-to-event endpoint, such as overall survival or progression-free survival, we assume that $X_i$ follows a Weibull distribution parameterized in terms of the median $\theta$ and the shape parameter $\kappa$: 
\begin{equation*}
X_i \sim \mbox{Weibull}(\theta,\kappa),
\end{equation*}
with probability density function
$
f(X_i|\theta,\kappa)=(ln2)\kappa\theta^{-\kappa}{X_i}^{\kappa-1}\mbox{exp}\big\{-(ln2)\theta^{-\kappa}{X_i}^{\kappa}\big\},
$
and survival function
$S(X_i|\theta,\kappa)=\mbox{exp}\big\{ -ln(2)\theta^{-\kappa}X_i^{\kappa} \big\}$.
This parameterization expresses the Weibull distribution in terms of the median $\theta$, which often has a more interpretable clinical interpretation, and the shape parameter $\kappa$. At any interim decision time, let $t_i$ denote the observed time to event or censoring time for the $i$th patient, and let $\zeta_i=I(t_i=X_i)$ denote the censoring indicator. The likelihood for patient $i$, based on data $(t_i,\zeta_i)$, is given by $L\big((t_i,\zeta_i)|\theta,\kappa\big)=f(t_i|\theta,\kappa)^{\zeta_i}S(t_i|\theta,\kappa)^{1-\zeta_i}$. Given interim data ${\cal D}_n=\big((t_1,\zeta_1),\cdots,(t_n,\zeta_n)\big)$ from the first $n$ patients, the joint likelihood function is $L({\cal D}_n|\theta,\kappa)=\prod_{i=1}^nL\big((t_i,\zeta_i)|\theta,\kappa\big)$.

We assume user-specified priors $\pi(\theta)=\mbox{Gamma}(\phi,\eta)$ and $\pi(\kappa)= \mbox{Gamma}(\lambda,\gamma)$  for the median $\theta$ and shape parameter $\kappa$, respectively. Given interim data ${\cal D}_n$, the joint posterior distribution of $(\theta,\kappa)$ under the user-specified prior is $\pi(\theta,\kappa|{\cal D}_n)\propto \pi(\theta)\pi(\kappa)L({\cal D}_n|\theta,\kappa)$, which can be sampled using the Gibbs sampler \citep{Geman84}\\[12pt]
\noindent\textit{RCT with time-to-event outcomes}: For an RCT with time-to-event outcomes, we assume a Cox proportional hazards model of the form
\begin{equation*}
h(t|\theta,\kappa,\beta)=h_0(t|\theta,\kappa)\mbox{exp}(\beta Z), \label{cox}
\end{equation*}
where $Z$ is the treatment indicator ($Z=1$ for the experimental treatment and $Z=0$ for the control), and  $\mbox{exp}(\beta)$ represents the hazard ratio (HR) comparing the experimental treatment to the control.

We specify the baseline hazard function $h_0(t|\theta,\kappa)=(ln2))\theta^{-\kappa}\kappa t^{\kappa-1}$, corresponding to a Weibull distribution parameterized by the median $\theta$ and shape parameter $\kappa$. The resulting probability density function and survival function are
$
f(t|\theta,\kappa,\beta)=(ln2)\kappa\theta^{-\kappa}{t}^{\kappa-1}\mbox{exp}(\beta Z)$
$\mbox{exp}(-(ln2)\theta^{-\kappa}\mbox{exp}(\beta Z){t}^{\kappa}),
$
and
$
S(t|\theta,\kappa,\beta)=\mbox{exp}\big( -ln(2)\theta^{-\kappa}\mbox{exp}(\beta Z)t^{\kappa} \big).
$
The likelihood for patient $i$ with data $(t_i,\zeta_i)$ is $L\big((t_i,\zeta_i)|\theta,\kappa,\beta\big)=f(t_i|\theta,\kappa,\beta)^{\zeta_i}S(t_i|\theta,\kappa,\beta)^{1-\zeta_i}$. Given interim data ${\cal D}_n=\big((t_1,\zeta_1),\cdots,(t_n,\zeta_n)\big)$ from the first $n$ patients, the joint likelihood function is $L({\cal D}_n|\theta,\kappa,\beta)=\prod_{i=1}^nL\big((t_i,\zeta_i)|\theta,\kappa,\beta\big)$. We assume user-specified priors $\pi(\theta)=\mbox{Gamma}(\phi,\eta)$, $\pi(\kappa)= \mbox{Gamma}(\lambda,\gamma)$, and $\pi(\beta)=\mbox{Normal}(\xi,\tau^2)$ for the parameters $\theta$, $\kappa$, and $\beta$, respectively. The joint posterior distribution of $(\theta,\kappa,\beta)$ under the user-specified prior is $\pi(\theta,\kappa,\beta|{\cal D}_n)\propto \pi(\theta)\pi(\kappa)\pi(\beta)L({\cal D}_n|\theta,\kappa,\beta)$, which can be sampled using the Gibbs sampler.

\subsection{Decision rule and inferential metrics}
Let $N$ denote the maximum sample size of the trial and $n$ an interim sample size. Let $\theta_0$ denote the treatment effect for the control. In single-arm trials, $\theta_0$ is assumed to be a known constant, whereas in two-arm RCTs, $\theta_0$ is treated as an unknown parameter.  The treatment is considered effective, relative to the control,  if $\theta > \theta_0$.  

We consider the following Bayesian superiority decision rule, evaluated at each interim analysis and, if the trial is not stopped early, again at the final analysis:
\begin{quote}
Stop the trial and claim that the treatment is effective if $\mbox{Pr}(\theta > \theta_0  \,|\, {\cal D}_n) > C$, where $C \in (0,1)$ is a probability cutoff. 
\end{quote}
For ease of exposition, we refer to the trial design with a fixed sample size $N$ and no interim analyses as the {\it fixed design}, and the design that incorporates interim analyses and allows early stopping as the {\it adaptive design}. For notational simplicity, we assume hereafter that  $\theta_0$ is a constant. However, all assessments and results presented later also apply to settings in which $\theta_0$ is random, as in RCTs.

To isolate the effect of interim analyses, we apply the same decision rule when evaluating the operating characteristics of both the fixed and adaptive designs. In other words, we do not calibrate the cutoff $C$ separately for the two designs. This ensures that any observed differences can be attributed solely to the presence of interim analyses.

In addition, in our assessment, we adhere to the fundamental assumption of the Bayesian framework that $\theta$ is random. This avoids the philosophical argument that evaluating Bayesian designs using the frequentist criteria (i.e., assuming $\theta$ is fixed, such as evaluating type I error at a fixed null value) is neither meaningful nor interpretable. This argument has been a main barrier to reaching a consensus on whether Bayesian methods can accommodate multiple interim looks without any consequences. 

Treating $\theta$ as a random variable necessitates distinguishing between the {\it data-generating prior} $\pi_g(\theta)$ and the {\it user-specified prior} $\pi(\theta)$ used for Bayesian inference. We assume that the data are generated according to the Bayesian process: sampling $\theta$ from $\pi_g(\theta)$, and generating data ${\cal D}_n$ based on $f(X_i|\theta)$.  $\pi_g(\theta)$ can be interpreted as the distribution of the treatment effect $\theta$ across the population of all trials within a therapeutic area, including positive and negative trials, which is generally unknown.  In practice, $\pi_g(\theta)$ is typically unknown. For a specific trial, users have to specify a prior to make posterior inferences and interim decisions. They assume a user-specified prior $\pi(\theta)$ (e.g., a noninformative prior) and, conditional on the observed data ${\cal D}_n$, calculate $\mbox{Pr}(\theta|{\cal D}_n)$ to make decisions.

We explore scenarios in which  $\pi(\theta)$ and $\pi_g(\theta)$ are either identical or different; in this paper, we refer to these cases as $\pi(\theta)$ being correctly specified or misspecified, respectively.  We assume that the data model $f(X_i|\theta)$ is always correctly specified to focus on the impact of interim analyses.  To evaluate the impact of interim analyses on trial operating characteristics, we consider the following key performance metrics:

\begin{itemize}

\item Bias of posterior mean: The expected difference between $\hat{\theta}=\mathbb{E}(\theta|{\cal D})$ and the true value of $\theta$, i.e., $\mathbb{E}_{\theta \sim \pi_g, \mathcal{D} \sim f(\cdot \mid \theta)}[\hat{\theta} - \theta]$. 
\item MSE of posterior mean: Defined as $\mathbb{E}_{\theta \sim \pi_g, \mathcal{D} \sim f(\cdot \mid \theta)}[(\hat{\theta} - \theta)^2]$.
\item Coverage of 95\% posterior credible intervals: The probability that a 95\% posterior credible interval of $\theta$ contains the true value of $\theta$, i.e., $\Pr_{\theta \sim \pi_g, \mathcal{D} \sim f(\cdot \mid \theta)}(\theta \in CI)$, where $CI$ represents a 95\% credible interval for \(\theta\) derived from the posterior distribution $\pi(\theta \mid \mathcal{D})$ under $\pi(\theta)$. Specifically, in our simulation, we consider symmetric credible intervals $(\theta_{0.025}, \theta_{0.975})$, where $\theta_{\alpha}$ represents the $\alpha$-th percentile of the posterior distribution $\pi(\theta \mid \mathcal{D})$.

\item FDR: The expected proportion of trials in which the treatment is actually not effective (i.e.,  $\theta \le \theta_0$) among all trials that declare the treatment effective.  Let \(R\) denote the number of trials that claim the treatment is effective (i.e., discoveries), and let \(V\) denote the number of false claims, where the treatment is in fact not effective.  We consider both the original FDR, introduced by \citet{Benjamini95}, and the positive FDR (pFDR), proposed by \citet{Storey03}. These are defined as  FDR$=\mathbb{E}_{\theta \sim \pi_g, \mathcal{D} \sim f(\cdot \mid \theta)}\left[\frac{V}{R}|R>0\right] \mbox{Pr}(R>0) = \mathbb{E}_{\theta \sim \pi_g, \mathcal{D} \sim f(\cdot \mid \theta)}\left[\frac{V}{R \vee 1}\right]$, where \( R \vee 1 = \max(R, 1) \) to avoid division by zero, and pFDR$=\mathbb{E}_{\theta \sim \pi_g, \mathcal{D} \sim f(\cdot \mid \theta)}\left[\frac{V}{R}|R>0\right]$, respectively. 

Although the FDR was originally developed as a tool to control frequentist error rates, \citet{Storey03} noted that it is inherently more Bayesian in nature, since it is defined conditional on the observed data, in contrast to the type I error rate, which is defined conditional on the true parameter  $\theta$. 

The FDR/pFDR is of substantial practical interest. If we assume that positive trials lead to drug approvals, then the pFDR quantifies, among all approval decisions, the proportion that are incorrect, that is, cases in which the approved drug is actually not effective. From this perspective, the pFDR can be interpreted as the risk of regulatory regret.

\item ATIE (Average Type I Error Rate): The probability of claiming the treatment effective when it is actually not effective (i.e., $\theta \le \theta_0 $). Formally, ATIE $ =\mathbb{E}_{\theta \sim \pi_g, \mathcal{D} \sim f(\cdot \mid \theta)} [ {\cal S} \,|\, \theta \le \theta_0] $, where ${\cal S}= I\{\mbox{Pr}(\theta > \theta_0 \,|\, {\cal D}_n) > C\}$. The ATIE can be regarded as the Bayesian counterpart of the frequentist type I error rate,  as it averages over the distribution of the $\theta$ in the region $\theta \le \theta_0$ (i.e., when the treatment is not effective), treating $\theta$ as a random variable rather than a fixed parameter. The ATIE is sometimes referred to as the Bayesian type I error rate \citep{best2025beyond}, and is described in the FDA’s Guidance for the Use of Bayesian Statistics in Medical Device Clinical Trials as an alternative definition of type I error for Bayesian methods \citep{FDA10}.

\end{itemize}
A few remarks are warranted. First, in the definition and calculation of the above performance metrics, $\theta$ is random.  This setup aims to preempt the argument that the evaluation is not consistent with the Bayesian framework. 
Second, our focus here is solely on evaluating the impact of repeated interim analyses on these commonly used performance metrics. It is not our goal to argue for or against the importance or appropriateness of any particular metric, especially from a philosophical standpoint (e.g., some may argue that ATIE conditions on $\theta$, whereas for a given trial Bayesians are primarily interested in conditioning on the observed data ${\cal D}$). Rather, users can decide for themselves how relevant a given metric is to their specific application.
Lastly, while we use the generic notation $\theta$, in the context of RCTs with time-to-event outcomes, the inferential quantities are computed with respect to the HR, $\exp({\beta})$.

\subsection{Impact of interim stopping \label{sec:impact}}
In this section, we examine the impact of incorporating interim analyses and early superiority stopping on various design performance metrics, using the corresponding Bayesian fixed design (with no interims) as the reference. 
In this way, the only distinction between the two designs lies in the inclusion of interim analyses.
This allows us to focus on whether incorporating interim analyses and stopping rules alters the operating characteristics of Bayesian inference, avoiding unnecessary debate over the frequentist versus Bayesian framework.  Table~\ref{impact} summarizes our findings, which are general and not specific to any particular model or the Bayesian decision rule discussed earlier. 

\subsection*{Bias of the posterior mean}

Let  $\mathcal{D}_n$ be the observed data at an interim stopping time $n <N$, governed by a pre-specified stopping rule (e.g., stop when $\mbox{Pr}(\theta>\theta_0 \, | \, {\cal D}_n) > C$), and let \(\hat{\theta}_n = \mathbb{E}_\pi [\theta \mid \mathcal{D}_n]\) denote the posterior mean estimates based on user-specified prior $\pi(\theta)$.  Under the data-generating model $\pi_g(\theta)$ and $f(\cdot \mid \theta)$, the bias of $\hat{\theta}_n$  is given by
\begin{eqnarray*}
\text{Bias}(\hat{\theta}_n) &=& \mathbb{E}_{\theta \sim \pi_g, \mathcal{D}_n \sim f(\cdot \mid \theta)}[\hat{\theta}_n] -\mathbb{E}_{\pi_g} [\theta] \\
&=& \mathbb{E}_{\pi_g} \left[ \mathbb{E}_{\mathcal{D}_n \mid \theta} \left[ \mathbb{E}_{\pi}[\theta \mid \mathcal{D}_n] \right] \right]
-\mathbb{E}_{\pi_g} [\theta].
\end{eqnarray*}

When \(\pi = \pi_g\), by the law of total expectation \citep{Casella02}, we have \[
\mathbb{E}_{\pi_g} \left[ \mathbb{E}_{\mathcal{D}_n \mid \theta} \left[ \mathbb{E}_{\pi}[\theta \mid \mathcal{D}_n] \right] \right] = 
\mathbb{E}_{\pi_g} \left[ \mathbb{E}_{\mathcal{D}_n \mid \theta} \left[ \mathbb{E}_{\pi_g}[\theta \mid \mathcal{D}_n] \right] \right]=
\mathbb{E}_{\mathcal{D}_n} \left[ \mathbb{E}_{\pi_g}[\theta \mid \mathcal{D}_n] \right] = \mathbb{E}_{\pi_g}[\theta].
\]
This  means that the posterior mean under $\pi_g$   is an unbiased estimator of the true treatment effect, even when $\mathcal{D}_n$  is selected adaptively through interim analyses. In other words, interim stopping does not introduce bias when the (user-specified) prior is correctly specified (i.e., \(\pi = \pi_g\)). In Bayesian terms, this is consistent with the likelihood principle: only the observed data $\mathcal{D}_n$  matters for inference, not the stopping rule. 

This result has been cited in some of the literature to support the claim that Bayesian analysis and stopping can be performed at any point during a trial without adjustment. However, what is often not explicitly highlighted is that this property holds only when the user-specified prior used for inference and decision-making is identical to the underlying data-generating prior. One goal of this article is to clarify this distinction.

Unfortunately, when \(\pi \ne \pi_g\), which is the dominant case in practice, $\mathbb{E}_{\pi_g} \left[ \mathbb{E}_{\mathcal{D}_n \mid \theta} \left[ \mathbb{E}_{\pi}[\theta \mid \mathcal{D}_n] \right] \right]$  generally \emph{does not} equal \( \mathbb{E}_{\pi_g}[\theta] \) because the outer expectation is under $\pi_g$, while the inner posterior is computed using $\pi$, and $ \mathbb{E}_{\pi}[\theta \mid \mathcal{D}_n] \ne \mathbb{E}_{\pi_g}[\theta \mid \mathcal{D}_n]$. Interim stopping selects favorable \( \mathcal{D}_n \) with higher-than-average treatment effects, which often amplifies prior misspecification, resulting in biased posterior mean estimate.

\subsection*{MSE of the posterior mean}

Under the generative model $\pi_g$ and $f(\cdot \mid \theta)$, the MSE of $\hat{\theta}_n$   is given by
\[
\text{MSE}(\hat{\theta}_n) = \mathbb{E}_{\theta \sim \pi_g, \mathcal{D}_n\sim f(\cdot \mid \theta)}[(\hat{\theta}_n - \theta)^2] = \text{Bias}^2(\hat{\theta}_n)+ \text{Var}(\hat{\theta}_n).
\]
As shown above, when \(\pi = \pi_g\), $\text{Bias}^2(\hat{\theta}_n)=0$. However, early stopping on average reduces sample size (i.e., $n<N$), which in turn increase posterior variance $\text{Var}(\hat{\theta}_n)$. Therefore, interim stopping increases MSE.

When \(\pi \ne \pi_g\), both bias and variance contribute to an even greater increase in the MSE of the posterior mean. This inflation is further amplified by interim stopping, as stopping rules tend to preferentially select data with extreme or favorable outcomes (e.g., early signs of efficacy), thereby skewing the posterior distribution toward overly optimistic estimates. The combination of prior misspecification and selection bias from early stopping leads to substantially degraded inferential accuracy.

\subsection*{Coverage of credible intervals}

Let \(CI_n\) denote a \(100(1-\alpha)\%\) Bayesian credible interval for \(\theta\) based on $\pi(\theta)$ and $\mathcal{D}_n$. The coverage of this interval is defined as $\Pr_{\theta \sim \pi_g, \mathcal{D}_n \sim f(\cdot \mid \theta)}(\theta \in CI_n).$

When \(\pi = \pi_g\), the posterior distribution is calibrated to the true data-generating process in the sense that 
\[
\Pr_{\theta \sim \pi_g, \mathcal{D}_n \sim f(\cdot \mid \theta)}(\theta \in CI_n) = 1 - \alpha,
\]
implying that Bayesian credible intervals maintain their nominal frequentist coverage, even under interim analyses with stopping rules \citep{Dawid82}. This result follows from the law of total expectations:
   \[
   \mathbb{E}_{\theta \sim \pi_g, \mathcal{D}_n \sim f(\cdot \mid \theta)}[\mathbf{1}_{\theta \in CI_n}]
   = \mathbb{E}_{\mathcal{D}_n}\left[ \mathbb{E}_{\theta \sim \pi_g(\cdot \mid \mathcal{D}_n)}[\mathbf{1}_{\theta \in CI_n}] \right],
   \]
 Since $\pi = \pi_g$, the posterior is calculated under the true prior $\pi_g$,  and by construction of the Bayesian credible interval, the inner expectation is:
   \[
   \mathbb{E}_{\theta \sim \pi_g(\cdot \mid \mathcal{D}_n)}[\mathbf{1}_{\theta \in CI_n}] = \Pr_{\theta\sim \pi_g} (\theta \in CI_n \mid \mathcal{D}_n) = 1 - \alpha.
   \]
  Hence,
   \[
   \Pr_{\theta, \mathcal{D}_n}(\theta \in CI_n) = \mathbb{E}_{\mathcal{D}_n}[1 - \alpha] = 1 - \alpha.
   \]
   
However, when \(\pi \ne \pi_g\), the posterior (and the resulting Bayesian credible interval) is misaligned with the data-generating process. As a result,  the credible interval may under- or over-cover the true parameter, leading to deviations from nominal coverage. This is because when $\pi \neq \pi_g$, the posterior used to construct credible interval is misspecified, and thus the inner expectation is:
   \[
   \mathbb{E}_{\theta \sim \pi_g(\cdot \mid \mathcal{D}_n)}[\mathbf{1}_{\theta \in CI_n}] \neq 1 - \alpha.
   \]
Therefore,
   \[
   \Pr_{\theta, \mathcal{D}_n}(\theta \in CI_n) = \mathbb{E}_{\mathcal{D}_n}[\mathbb{E}_{\theta \sim \pi_g(\cdot \mid \mathcal{D}_n)}[\mathbf{1}_{\theta \in CI_n}]] \neq 1 - \alpha.
   \]
This mismatch can be further amplified by interim stopping, as stopping rules tend to preferentially select data with extreme or favorable outcomes, which biases the posterior toward an overly optimistic direction. As a result, interim stopping can exacerbate the undercoverage of credible intervals when the prior is misspecified.

This result further demonstrates that the claim that Bayesian analysis and stopping can be performed at any point during a trial without adjustment is neither entirely true nor entirely false. It holds only when the user-specified prior used for inference and decision-making is identical to the underlying data-generating prior, a condition that rarely, if ever, holds in practice. This finding reconciles the conflicting arguments regarding Bayesian multiplicity that have long persisted in the literature.

\subsection*{FDR/pFDR}
Bayesian interim analyses and stopping generally lead to inflation of the FDR and pFDR for the following reason. Recall that the FDR is defined as  $\text{FDR} = \mathbb{E}\left[\frac{V}{R \vee 1}\right].$ Under a fixed rejection threshold $C$, conducting multiple interim analyses increases the likelihood of $\mbox{Pr}(\theta > \theta_0  \,|\, {\cal D}_n) > C$ due to random data fluctuations, particularly early in the trial. This leads to an accumulation of false  discovery across interim looks, widely known as multiplicity. In contrast, the number of true discovery when the treatment is truly effective tends to grow more slowly or even plateau. This is because the statistical power is typically high by design (e.g., 0.8), and additional interim looks offer diminishing returns in detecting true effects. As a result, the expected number of false discoveries ($V$) tends to increase more rapidly than the total number of discoveries ($R$), leading to inflation of the FDR (see Supplementary Materials for a proof). This inflation occurs even when the prior distribution is correctly specified (i.e., \(\pi = \pi_g\)), and may be further exacerbated under prior misspecification (i.e., \(\pi \ne \pi_g\)). A similar argument applies to the pFDR as well.

\subsection*{ATIE}

Bayesian interim analyses inflate ATIE, regardless of whether $\pi(\theta)$ is correctly or incorrectly specified. This is because at each interim analysis, there is a non-zero probability that the posterior probability exceeds the threshold, i.e., \(\Pr(\theta > \theta_0 \mid \mathcal{D}_n) > C\), purely due to random fluctuations in the observed data, when \(\theta \le \theta_0\). This resulting in inflation of ATIE. This inflation is often further exacerbated under prior misspecification (i.e., \(\pi \ne \pi_g\)), as stopping rules tend to preferentially select data with extreme or favorable outcomes (e.g., indicating early efficacy), thereby biasing the posterior toward an overly optimistic direction.

\section{Simulation}
\subsection{Simulation setting}
We conducted  simulation studies to evaluate the impact of repeated Bayesian interim analyses on various design performance metrics. The maximum sample size was set to $N=100$. The data-generating prior, $\pi_g(\theta)$, was assumed to belong to the same probability distribution family as  the user-specified prior, $\pi(\theta)$, but with potentially different hyperparameter values.  For each outcome type and each scenario defined by $\pi(\theta)$, we simulated 50,000 trials. In the $k$th simulated trial, we first sampled $\theta^k$ from $\pi_g(\theta)$, then generated interim data ${\cal D}^k_{n}=\{X^k_{1},\cdots, X^k_{n} \}$ from $f(X_i |\theta^k)$, e.g., $\mbox{Bernoulli}(\theta^k)$ for binary endpoints or $N(\theta^k,\sigma^2)$ for normal endpoints. We implemented two interim analyses at the 40th and 70th patients, and a final analysis at the 100th patient, i.e., for $n=40,\ 70$, and $100$,  the trial was stopped and claim that the treatment is effective if $\mbox{Pr}(\theta > \theta_0 |{\cal D}^k_{n})>C$.

As a benchmark, we also evaluated the same performance metrics for trials conducted without interim analyses (i.e., fixed design). In these trials, the decision of claiming that the treatment is effective was made at the end of the trial based on the criterion $\mbox{Pr}(\theta > \theta_0|{\cal D}^k_{N})>C$. The probability cutoff $C$ was calibrated to control the pFDR at 5\% under the scenario where $\pi(\theta)= \pi_g(\theta)$. This same cutoff was then applied to designs with and without interim analyses; therefore, any differences observed can be directly attributed to the inclusion of interim analyses. It is important to note that we did not re-calibrate the decision cutoff $C$ for each individual scenario. As a result, in scenarios where $\pi(\theta) \neq \pi_g(\theta)$, the pFDR and other metrics for the fixed design may deviate from the nominal level of 0.05. However, this does not pose a concern for our investigation, as the primary objective is to compare results with and without including interim analyses to assess their impact on operating characteristics, rather than to evaluate the absolute operating characteristics of a design across different scenarios.

Below are the simulation parameters for single-arm trials; those for RCTs are provided in Supplementary Materials. For binary endpoint, we set the response rate of the (historical) control as $\theta_0=0.6$ and assumed the true response rate of the experimental arm generated from  $\pi_g(\theta) = $ Beta(3, 3). We evaluated three classes of user-specified prior $\pi(\theta)$: (i) neutral priors centered at $\theta_0=0.6$; (ii) pessimistic priors centered at $0.5$; and (iii) optimistic priors centered at $0.7$. Within each class, we examined three priors with increasing spread (i.e., decreasing informativeness). Specifically, for $\pi(\theta)$ centered at 0.6, we considered Beta(18, 12), Beta(3.6, 2.4), and Beta(0.06, 0.04).  For $\pi(\theta)$ centered at 0.5, we considered Beta(15, 15), Beta(3, 3), and Beta(0.05, 0.05). Notably, Beta(3, 3) matches $\pi_g(\theta)$. And for $\pi(\theta)$ centered at 0.7, we considered Beta(21, 9), Beta(4.2, 1.8), and Beta(0.07, 0.03). The probability cutoff was calibrated as previously described, resulting in $C=0.689$. Due to the discrete nature of the observed data, it was not always possible to find a cutoff yielding an exact 5\% pFDR. This value of $C$ was selected to achieve an pFDR closest to, but not exceeding, 5\%.

For the normal outcome with $\sigma^2=1$, we assumed $\pi_g(\theta) =  N(0,1)$ and set $\theta_0=0.25$. The neutral priors $\pi(\theta)$ centered at 0.25 included $N(0.25, 0.2), N(0.25, 1)$, and $N(0.25, 1000)$. Optimistic and pessimistic priors were centered at 0.5 and 0, respectively. The probability cutoff was calibrated and set to $C=0.55$. For the normal outcome with unknown variance, we used the conjugate prior: $\pi_g(\theta|\sigma^2)=N(0,\sigma^2/5)$ and $\pi_g(\sigma^2)= \mbox{Inv-}\chi^2(5,40)$, with $\theta_0=0.25$. Priors $\pi(\theta,\sigma^2)$ were constructed with $\theta$ centered at 0.25 (neutral), 0 (pessimistic), and 0.5 (optimistic), as detailed in Table 4.   The probability cutoff was set at $C=0.63$. 

For the time-to-event outcomes, we assumed a 12-month follow-up period with patient accrual following a Poisson process at a rate of six patients per month. 
We set $\theta_0=8$ and used $\pi_g(\theta)=\mbox{Gamma}(12,2)$ and $\pi_g(\kappa)=\mbox{Gamma}(8,2)$. We evaluated three $\pi(\theta,\kappa)$ priors with $\theta$ centered at 8 (neutral), 6 (pessimistic), and 10 (optimistic), each with varying  precision (see Table 5). The cutoff was calibrated and set to $C=0.64$ to control the pFDR at 5\% in the absence of an interim analysis. 

The performance  metrics were calculated as follows. pFDR was calculated as the proportion of trials in which the treatment is not effective (i.e., $\theta\le\theta_0$) among those claiming that the treatment is effective, across the 50,000 simulated trials. FDR was then obtained by multiplying the pFDR by the proportion of trials claiming that the treatment is effective. Bias was determined as the average difference $\hat{\theta}^k-\theta^k$, and MSE was computed as the mean of $(\hat{\theta}^k-\theta^k)^2$, where $\hat{\theta}^k$ denotes the posterior mean of $\theta$ in the $k$th simulated trial. Coverage was defined as the percentage of trials in which $\theta^k$ fell within the 95\% posterior credible interval for $\theta$ in the $k$th trial. FPR was computed as the percentage of trials claiming that the treatment is effective among those in which the treatment is actually not effective with $\theta\le\theta_0$.

\subsection{Results}
Table 2 presents results for single-arm trials with a binary endpoint, using the fixed design without interim analyses as the reference. Unless otherwise noted, the discussion focuses on comparing designs with interim analyses to the corresponding fixed design. This allows us to isolate and assess the impact of conducting multiple interim looks, rather than comparing absolute operating characteristics across different scenarios. Our findings are consistent with the theoretical investigation in Section \ref{sec:impact}, indicating that interim analyses can have a substantial effect on the operating characteristics of the design.

In the ideal scenario where the user-specified prior matches the data-generating prior  (i.e., $\pi(\theta)=\pi_g(\theta)=$Beta(3, 3)), adding interim analyses does not affect the coverage rates of 95\% credible intervals, or the bias. However, incorporating interim analyses lead to inflation in pFDR, FDR, ATIE, and MSE. For example, incorporating interim analyses increased the pFDR from 0.046 to 0.095, FDR from 0.012 to 0.028, ATIE from 0.018 to 0.041, and MSE from 0.002 to 0.003.

When the user-specified prior differs from the data-generating prior (i.e., $\pi(\theta) \neq \pi_g(\theta)$), as is the case with pessimistic, optimistic, or neutral priors that do not match $\pi_g(\theta)$, adding interim analyses affects all operating metrics, leading to increases in pFDR, FDR, ATIE, and MSE, as well as changes in the coverage rates of credible intervals and bias. In particular, as $\pi(\theta)$ moves away from $\pi_g(\theta)$, coverage of the credible intervals may drop below nominal value. This is particularly concerning because, in practice, the user-specified prior is almost always different from the true data-generating prior. In fact, vague or noninformative priors are among the most commonly used. For example, when a noninformative prior Beta(0.05, 0.05) was used as the user-specified prior, incorporating interim analyses increased the pFDR from 0.060 to 0.131, ATIE from 0.025 to 0.062, bias from -0.3$\times 10^{-3}$ to 5.3$\times 10^{-3}$, and MSE from 0.002 to 0.003, while reducing the coverage of the credible interval from 0.948 to 0.944.

Tables 3 and 4 present analogous results for normal endpoints with known and unknown variances, and Table 5 shows findings for time-to-event endpoints. Across all endpoint types, patterns were broadly consistent with those described above for the binary case. Results for RCTs are generally similar to those of single-arm trials; see Tables S1-S4 in Supplementary Materials for more details.

\subsection{Sensitivity analyses}
We examined the effect of altering the frequency of interim analyses. For binary and normal outcomes, where patient-level data are typically observed promptly, we employed a fully sequential design in place of the original two interim analyses at the 40th and 70th patients. In this design, early stopping was assessed after each patient beginning with the 20th, i.e., for $n=20,\cdots, N$. For time-to-event endpoints, where outcomes are event-driven rather than patient-driven, interim analyses were conducted every 5 patients starting from the 20th, to allow time for meaningful accumulation of event information. Results for each outcome type under these settings are shown in Tables S5-S7 and were broadly consistent with those observed under the original design.

We further evaluated the performance metrics under alternative values of the probability that a treatment is truly effective, denoted by $\mbox{Pr}_g({\rm effective}) = \Pr_{\theta\sim \pi_g} (\theta>\theta_0)$, as determined by the data-generating prior distribution of $\theta$. In the context of drug development, $\mbox{Pr}_g({\rm effective})$ represents the proportion of all investigational drugs that are genuinely effective. For example, under a normal outcome with known variance 1 and an equivalence margin $\delta=0$, the prior $\pi_g(\theta)=N(0,1)$ yields $\mbox{Pr}_g({\rm effective})=0.4$. We also considered a more concentrated prior, $\pi_g(\theta)=N(0, 0.09)$, which reduces $\mbox{Pr}_g({\rm effective})$ to 0.2. 
The corresponding results for continuous, binary, and time-to-event outcomes in single-arm trials, presented in Tables S8 - S10 of the Supplementary Materials,  exhibited patterns consistent with those observed under the original prior specification. 

For the normal outcome with unknown variance, the scale parameter $\sigma_0^2$ in $\pi(\sigma^2)$ was originally set to 40, matching that in $\pi_g(\sigma^2)$. In sensitivity analyses, we varied $\sigma^2_0$ to 20 and 60. The resulting metrics, shown in Tables S11 and S12, remained consistent with those from the main setting, further supporting the robustness of the findings.

\section{Conclusion}
We have conducted a comprehensive investigation into the impact of repeated Bayesian interim analyses on key design metrics, including the bias and MSE of the posterior mean, the coverage probability of posterior credible intervals, FDR/pFDR, and ATIE. Our results show that the claim—that Bayesian interim analyses with superiority stopping are free from multiplicity concerns and may be performed at any time without adjustment—holds for the posterior mean and credible-interval coverage only when the user-specified prior coincides exactly with the true data-generating prior, an assumption that is rarely, if even, satisfied in practice. Moreover, we show that even under this ideal scenario, repeated interim analyses can still systematically distort FDR, pFDR, and ATIE. More importantly, when the user-specified prior differs from the true data-generating prior, which is typically the case in real-world applications, all of these metrics can be adversely affected. These theoretical findings are supported by extensive simulation studies, which confirm that repeated interim analyses can substantially alter the operating characteristics of Bayesian trials, especially when vague or noninformative priors are used. This is particularly concerning because such priors are commonly employed in practice.
 
These findings highlight the importance of accounting for the cumulative impact of repeated Bayesian interim decisions on the posterior distribution and subsequent inference. To safeguard against erroneous inference, researchers should consider strategies such as calibrating the decision threshold $C$ to account for the impact of interim analyses. This adjustment parallels multiplicity control methods in the frequentist paradigm, such as adopting stricter significance thresholds. An alternative approach is to specify a conservative, informative user-specified prior $\pi(\theta)$ that offsets the impact of interim analyses without requiring adjustment to $C$. However, this approach may be challenging to interpret. Equally critical is the need for thoughtful prior elicitation and comprehensive sensitivity analyses to evaluate the impact of prior misspecification. While Bayesian methods offer powerful tools for flexible and adaptive decision-making, their reliability hinges on a clear understanding of how repeated analyses interact with prior assumptions and inferential goals.

\bigskip

\noindent {\Large\bf Disclaimer} \\
This article reflects the views of the author and should not be construed to represent FDA's views or policies.


\bibliographystyle{apacite}
\bibliography{BayIntermRef}
\clearpage

\begin{table}[ht]
\centering
\caption{Impact of interim superiority stopping on design performance metrics, relative to the fixed design without interim analyses, assuming the same decision rule for declaring effectiveness.  \label{impact}}
\begin{tabular}{|p{2cm}|p{2cm}|p{4.5cm}|p{2cm}|p{4.5cm}|}
\hline
\textbf{Metric} & \textbf{Including interims when $\pi = \pi_g$} & \textbf{Justification} & \textbf{Including interims when $\pi \ne \pi_g$} & \textbf{Justification} \\
\hline
Posterior mean bias & Maintains unbiased & Follows from the law of total expectation under the correct prior. &  Biased & Prior misspecification combined with selective stopping distorts the posterior mean. \\
\hline
Credible interval coverage & Maintains nominal level & Credible intervals are calibrated under the prior predictive distribution. & Degraded (over or undercoverage) & Misspecified prior yields miscalibrated posteriors. \\
\hline
Posterior mean MSE & Increased & Early stopping raises variance due to smaller sample size. & Often further increased$^*$ & Both bias (from prior) and variance (from stopping) contribute to MSE inflation. \\
\hline
FDR/pFDR & Increased & Multiple interim looks increase the chance of stopping on spurious signals when the treatment actually is not effective. & Often further increased$^*$ & Prior misspecification distorts posterior probabilities, often worsening FDR and pFDR. \\
\hline
ATIE & Increased & More looks at the data raise the probability of false positives. & Often further increased$^*$ & Same mechanism, compounded by posterior miscalibration. \\
\hline
\end{tabular}
$^*$ Not necessarily increased, as it depends on whether $\pi$ is optimistic or pessimistic on efficacy relative to $\pi_g$.
\end{table}

\begin{table}
\begin{center}
\begin{footnotesize}
\caption{Single-arm trials with binary endpoints: pFDR, FDR, bias, MSE,  coverage of posterior credible intervals, and ATIE, evaluated with or without interim stopping, under various user-specified priors with $\theta_0=0.6$ and data-generating prior $\pi_g(\theta) = \mbox{Beta}(3,3)$. }
\begin{tabular}{lccccccc}
\hline
\hline

$\pi(\theta)$ &      Interims &        pFDR &     FDR &     Bias  ($\times 10^{-3}$)                    &    MSE & Coverage &       ATIE     \\

\hline

Beta(18, 12) &         no &    0.060&0.017&23.0&0.004&0.793&0.025 \\

           &        yes &     0.123&0.039&19.1&0.005&0.777&0.057\\

           &            &            &            &            &            &            &                   \\

Beta(3.6, 2.4) &         no &   0.060&0.017&5.4&0.002&0.949&0.025 \\

           &        yes &      0.123&0.039&7.7&0.003&0.948&0.057 \\

          &            &            &            &            &            &            &                   \\

Beta(0.06, 0.04) &         no &      0.060&0.017&-0.2&0.002&0.948&0.025\\

           &        yes &     0.131&0.042&5.4&0.003&0.944&0.062 \\
           
          &            &            &            &            &            &            &                   \\

Beta(15, 15) &         no &    0.024&0.006&-0.1&0.003&0.822&0.008 \\

           &        yes &   0.038&0.009&-7.40&0.0&0.798&0.014 \\

              &            &            &            &            &            &            &                   \\

Beta(3, 3)$^*$ &         no &     0.046&0.012&-0.3&0.002&0.950&0.018\\

           &        yes &      0.095&0.028&-0.1&0.003&0.949&0.041\\

                &            &            &            &            &            &            &                   \\

Beta(0.05, 0.05) &         no &  0.060&0.017&-0.3&0.002&0.948&0.025 \\

           &        yes &     0.131&0.042&5.3&0.003&0.944&0.062 \\

                &            &            &            &            &            &            &                   \\

Beta(21, 9) &         no &   0.122&0.040&46.0&0.005&0.713&0.059 \\

           &        yes &    0.265&0.109&50.9&0.007&0.692&0.160 \\

                     &            &            &            &            &            &            &                   \\

Beta(4.2, 1.8) &         no &    0.079&0.023&11.1&0.002&0.943&0.034 \\

           &        yes &   0.166&0.057&16.0&0.003&0.940&0.084 \\

                     &            &            &            &            &            &            &                   \\
                     
Beta(0.07, 0.03) &         no &     0.060&0.017&-0.1&0.002&0.948&0.025\\

           &        yes &     0.131&0.042&5.6&0.003&0.944&0.062 \\

\hline
\hline
\end{tabular}

$^*$User-specified prior matches the data-generating prior\\
\end{footnotesize}
\end{center}
\end{table}

\begin{table}
\begin{center}
\begin{footnotesize}
\caption{Single-arm trials with normal endpoints with a known variance of 1: pFDR, FDR, bias, MSE,  coverage of posterior credible intervals, and ATIE, evaluated with or without interim stopping, under various user-specified priors with $\theta_0=0.25$ and data-generating prior $\pi_g(\theta) = N(0,1)$. }
\begin{tabular}{lccccccc}
\hline
\hline

$\pi(\theta)$ &      Interims &        pFDR &     FDR &     Bias  ($\times 10^{-3}$)                       &    MSE & Coverage &       ATIE    \\

\hline

N(0.25, 0.2) &         no &      0.051&0.021&12.2&0.012&0.925&0.035 \\

           &        yes &      0.098&0.043&-0.8&0.020&0.915&0.072 \\

                  &            &            &            &            &            &            &                   \\

N(0.25, 1) &         no &    0.051&0.021&2.7&0.010&0.949&0.035 \\

           &        yes &     0.098&0.043&4.2&0.016&0.949&0.072 \\

                      &            &            &            &            &            &            &                   \\

N(0.25, 1000) &         no &      0.051&0.021&0.2&0.010&0.950&0.035 \\

           &        yes &      0.098&0.043&6.0&0.017&0.950&0.072 \\

                &            &            &            &            &            &            &                   \\

 N(0, 0.2) &         no &     0.045&0.018&0.3&0.011&0.927&0.030 \\

           &        yes &      0.084&0.036&-19.3&0.021&0.912&0.060 \\

                     &            &            &            &            &            &            &                   \\

   N(0, 1)$^*$ &         no &      0.050&0.020&0.2&0.010&0.950&0.034 \\

           &        yes &    0.095&0.041&0.3&0.016&0.949&0.069 \\

                &            &            &            &            &            &            &                   \\

N(0, 1000) &         no &     0.051&0.021&0.2&0.010&0.950&0.035 \\

           &        yes &    0.098&0.043&6.0&0.017&0.950&0.072\\

                  &            &            &            &            &            &            &                   \\

N(0.5, 0.2) &         no &     0.058&0.024&24.1&0.012&0.919&0.040 \\

           &        yes &     0.113&0.051&18.1&0.020&0.911&0.085 \\

                  &            &            &            &            &            &            &                   \\

 N(0.5, 1) &         no &    0.053&0.022&5.1&0.010&0.949&0.036\\

           &        yes &       0.100&0.044&8.2&0.016&0.949&0.074 \\

                   &            &            &            &            &            &            &                   \\

N(0.5, 1000) &         no &      0.051&0.021&0.2&0.010&0.950&0.035 \\

           &        yes & 0.098&0.043&6.0&0.017&0.950&0.072 \\
           
\hline
\hline
\end{tabular}

$^*$User-specified prior matches the data-generating prior\\
\end{footnotesize}
\end{center}
\end{table}

\begin{table}
\begin{center}
\begin{footnotesize}
\caption{Single-arm trials with normal endpoints with an unknown variance: pFDR, FDR, bias, MSE,  coverage of posterior credible intervals, and ATIE, evaluated with or without interim stopping, under various user-specified priors with $\theta_0=0.25$ and data-generating prior $\pi_g(\theta|\sigma^2)= N(0,\sigma^2/5)$ and $\pi_g(\sigma^2)= \mbox{Inv-}\chi^2(5,40)$. }
\begin{tabular}{lccccccc}
\hline
\hline

$\pi(\theta)$ &      Interims &        pFDR &     FDR &     Bias    ($\times 10^{-3}$)                    &    MSE & Coverage &       ATIE    \\

\hline

$\theta|\sigma^2\sim N(0.25,\sigma^2/20)$ &         no &   0.050&0.022&39.4&0.837&0.895&0.041 \\

$\sigma^2\sim Inv-\chi^2(20,40)$ &        yes &     0.094&0.045&-103.4&1.465&0.872&0.085 \\

           &            &            &            &            &            &            &                   \\

$\theta|\sigma^2\sim N(0.25,\sigma^2/5)$ &         no &      0.051&0.023&11.6&0.633&0.950&0.043 \\

$\sigma^2\sim Inv-\chi^2(5,40)$ &        yes &      0.099&0.048&15.4&1.017&0.950&0.090 \\

           &            &            &            &            &            &            &                   \\

$\theta|\sigma^2\sim N(0.25,\sigma^2/0.1)$ &         no &      0.052&0.023&0.7&0.664&0.950&0.043 \\

$\sigma^2\sim Inv-\chi^2(0.1,40)$ &        yes &   0.100&0.049&90.5&1.119&0.949&0.092 \\

           &            &            &            &            &            &            &                   \\
$\theta|\sigma^2\sim N(0,\sigma^2/20)$ &         no &    0.046&0.020&-2.3&0.836&0.896&0.038 \\

$\sigma^2\sim Inv-\chi^2(20,40)$ &        yes &      0.084&0.039&-162.1&1.494&0.871&0.074\\

           &            &            &            &            &            &            &                   \\

$\theta|\sigma^2\sim N(0,\sigma^2/5)^*$ &         no &    0.050&0.022&-0.4&0.633&0.950&0.041 \\

$\sigma^2\sim Inv-\chi^2(5,40)^*$ &        yes &     0.095&0.046&-3.6&1.013&0.950&0.086 \\

           &            &            &            &            &            &            &                   \\
$\theta|\sigma^2\sim N(0,\sigma^2/0.1)$ &         no &     0.052&0.023&0.4&0.664&0.950&0.043 \\

$\sigma^2\sim Inv-\chi^2(0.1,40)$ &        yes &      0.100&0.049&90.0&1.119&0.949&0.092 \\

           &            &            &            &            &            &            &                   \\

$\theta|\sigma^2\sim N(0.5,\sigma^2/20)$ &         no &     0.055&0.025&81.0&0.842&0.893&0.046\\

$\sigma^2\sim Inv-\chi^2(20,40)$ &        yes &      0.106&0.052&-43.2&1.447&0.873&0.098 \\

           &            &            &            &            &            &            &                   \\
$\theta|\sigma^2\sim N(0.5,\sigma^2/5)$ &         no &    0.053&0.023&23.5&0.634&0.950&0.044 \\

$\sigma^2\sim Inv-\chi^2(5,40)$ &        yes &      0.102&0.049&34.4&1.020&0.950&0.093 \\

           &            &            &            &            &            &            &                   \\
           
$\theta|\sigma^2\sim N(0.5,\sigma^2/0.1)$ &         no &      0.052&0.023&0.9&0.664&0.950&0.043 \\

$\sigma^2\sim Inv-\chi^2(0.1,40)$ &        yes &    0.100&0.049&90.8&1.119&0.949&0.092 \\

\hline
\hline
\end{tabular}

$^*$User-specified prior matches the data-generating prior\\
\end{footnotesize}
\end{center}
\end{table}

\begin{table}
\begin{center}
\begin{footnotesize}
\caption{Single-arm trials with time-to-event endpoints: pFDR, FDR, bias, MSE,  coverage of posterior credible intervals, and ATIE, evaluated with or without interim stopping, under various user-specified priors with $\theta_0=8$ and data-generating prior $\pi_g(\theta)=\mbox{Gamma}(12,2)$ and $\pi_g(\kappa)=\mbox{Gamma}(8,2)$.}
\begin{tabular}{lccccccc}
\hline
\hline

$\pi(\theta)$ &      Interims &        pFDR &     FDR &     Bias  ($\times 10^{-3}$)                    &    MSE & Coverage &       ATIE    \\

\hline

$\theta\sim\mbox{Gamma}(128,16)$ &         no &      0.072&0.009&166.5&0.121&0.739&0.010 \\

$\kappa\sim\mbox{Gamma}(32,8)$ &        yes &      0.160&0.022&134.5&0.215&0.725&0.025 \\

           &            &            &            &            &            &            &                   \\

$\theta\sim\mbox{Gamma}(21.33, 2.67)$ &         no &     0.061&0.007&36.0&0.051&0.942&0.008\\

$\kappa\sim\mbox{Gamma}(8, 2)$ &        yes &      0.211&0.031&59.0&0.155&0.939&0.036 \\

           &            &            &            &            &            &            &                   \\

$\theta\sim\mbox{Gamma}(0.64, 0.08)$ &         no &      0.058&0.007&-0.6&0.051&0.948&0.008\\

$\kappa\sim\mbox{Gamma}(0.16, 0.04)$ &        yes &      0.357&0.066&538.7&3.985&0.936&0.075\\

           &            &            &            &            &            &            &                   \\

$\theta\sim\mbox{Gamma}(72, 12)$ &         no &     0.017&0.002&-15.8&0.072&0.892&0.002\\

$\kappa\sim\mbox{Gamma}(32,8)$ &        yes &   0.022&0.002&-53.3&0.153&0.877&0.003\\

             &            &            &            &            &            &            &                   \\

$\theta\sim\mbox{Gamma}(12, 2)^*$ &         no &     0.050&0.006&-1.6&0.049&0.949&0.007\\

$\kappa\sim\mbox{Gamma}(8, 2)^*$ &        yes &     0.116&0.015&-5.7&0.104&0.948&0.017\\

             &            &            &            &            &            &            &                   \\

$\theta\sim\mbox{Gamma}(0.36, 0.06)$ &         no &     0.059&0.007&-1.8&0.051&0.949&0.008\\

$\kappa\sim\mbox{Gamma}(0.16, 0.04)$ &        yes &     0.351&0.064&572.8&4.867&0.936&0.073 \\

             &            &            &            &            &            &            &                   \\
$\theta\sim\mbox{Gamma}(200, 20)$ &         no &       0.235&0.038&428.4&0.362&0.420&0.043\\

$\kappa\sim\mbox{Gamma}(32, 8)$ &        yes &      0.859&0.770&2759.4&10.062&0.122&0.882\\

             &            &            &            &            &            &            &                   \\

$\theta\sim\mbox{Gamma}(33.33, 3.33)$ &         no &     0.082&0.010&91.9&0.064&0.914&0.012\\

$\kappa\sim\mbox{Gamma}(8, 2)$ &        yes &      0.571&0.164&476.7&1.090&0.858&0.188 \\

               &            &            &            &            &            &            &                   \\

$\theta\sim\mbox{Gamma}(1, 0.1)$ &         no &      0.059&0.007&1.2&0.051&0.949&0.008 \\

$\kappa\sim\mbox{Gamma}(0.16, 0.04)$ &        yes &     0.368&0.069&550.3&3.772&0.937&0.079\\

\hline
\hline
\end{tabular}

$^*$User-specified prior matches the data-generating prior \\
\end{footnotesize}
\end{center}
\end{table}

\end{document}